\documentclass[12pt]{article}
\usepackage{epsfig}

\parskip=\medskipamount 
plus.3em minus.5em \abovedisplayshortskip=.5em plus.2em minus.4em
\renewcommand{\thesection}{\arabic{section}}

\catcode`@=11

\@addtoreset{equation}{section} \@addtoreset{equation}{subsection}
\def\theequation{\ifnum\value{section}=0 \arabic{equation}\ignorespaces
\else \ifnum\value{section}=-1 A.\arabic{equation}\ignorespaces
\else \ifnum\value{subsection}=0
\thesection.\arabic{equation}\ignorespaces \else
\thesection.\arabic{subsection}.\arabic{equation}\ignorespaces
                             \fi
                        \fi
                   \fi}

{\catcode`\'=\active \def'{{}^\bgroup\prim@s}}

\catcode`@=12

\newcommand{\bq}{\begin{equation}}
\newcommand{\be}{\begin{equation}}
\newcommand{\fq}{\end{equation}}
\newcommand{\ee}{\end{equation}}
\newcommand{\bqr}{\begin{eqnarray}}
\newcommand{\beqs}{\begin{eqnarray}}
\newcommand{\fqr}{\end{eqnarray}}
\newcommand{\eeqs}{\end{eqnarray}}

\newcommand{\rf}[1]{(\ref{#1})}

\def\bop#1{\setbox0=\hbox{$#1M$}\mkern1.5mu
    \vbox{\hrule height0pt depth.04\ht0
    \hbox{\vrule width.04\ht0 height.9\ht0 \kern.9\ht0
    \vrule width.04\ht0}\hrule height.04\ht0}\mkern1.5mu}

\begin{document}
\thispagestyle{empty}

\begin{flushright}
\begin{tabular}{l}
hep-th/0509027 \\
\end{tabular}
\end{flushright}

\vskip .6in
\begin{center}

{\bf Explicit Generation of Integer Solutions via CY manifolds}

\vskip .6in

{\bf Gordon Chalmers}
\\[5mm]

{e-mail: gordon@quartz.shango.com}

\vskip .5in minus .2in

{\bf Abstract} 

\end{center}

Metrics on Calabi-Yau manifolds are used to derive a formula that 
finds the existence of integer solutions to polynomials.  These 
metrics are derived from an associated algebraic curve, together 
with its anti-holomorphic counterpart.  The integer points in the 
curve coincide with points on the manifold, and the metric form 
around these points are used to find their existence.  The explicit 
form of the metrics can be found through a solution to the D-terms 
in a non-linear sigma model.

\vfill\break

\vskip .2in

The metrics on general CY manifolds have recently been computed in 
\cite{Chalmers1},\cite{Chalmers2}.  This 
task was accomplished by using a field theoretic count of classical tree 
graphs in scalar field theories\footnote{A subtlety associated with a 
deformation parameter can be circumvented with a variant of the procedure 
involving background $\phi$ line dependence.}  In prior work, the metrics 
of these 
Calabi-Yau manifolds was shown to provide means to finding both solutions 
to systems of algebraic equations and non-linear partial differential 
equations \cite{Chalmers3},\cite{Chalmers4} (with related work in 
\cite{Chalmers5}-\cite{Chalmers7}).  In this work explicit 
formulae are given that generate the 
integer solutions to polynomial equations using the metric form of these 
manifolds.  

\vskip .2in 
\noindent{\it Metric Expansions and Integer Solutions}

The starting point is the algebraic equation 

\bqr 
\sum a_{\sigma} \prod z_{\sigma(i)}^{\rho(i)} = 0 \ , 
\label{polynomial}
\fqr 
with $\rho(i)$ labeling the exponent of the coordinate $z_{\sigma(i)}$, whose 
index is labeled by the set $\sigma(i)$.   To every one of these equations, 
or sets of these equations, there is a Calabi-Yau metric.  These metrics can 
be formulated by a quotient construction using for example a non-linear 
${\cal N}=2$ sigma model.  The D-term solution of these models generates 
the K\"ahler potential, and the D-terms were solved in \cite{Chalmers2}.  

The K\"ahler potential on these metrics, in the patch containing the 
coordinates at $\phi_i=0$ has the expansion, 

\bqr 
{\cal K}= \sum b_\omega \prod \phi_{\omega(i)} \ .  
\label{Kahler}
\fqr  
Due to the polynomial form in \rf{polynomial}, there is a resovable 
singularity at the origin at $\phi_i=0$ in the metric derived from the 
K\"ahler potential.  This is not apparent in the solution obtained to 
\rf{Kahler} due to the classical field tree diagram count in 
\cite{Chalmers1},\cite{Chalmers2}, but it can be obtained by 
resumming the series with the numbers $b_\omega$.  The resummations of the 
these $b_\omega$ seems complicated, but there is an alternative to obtaining 
the analytic continuation that extracts the branch cuts.  

The expansion to the metric about any integer sets $z_i=p_i$ to the 
equation in 
\rf{polynomial} can be obtained by following the same classical graph count in 
\cite{Chalmers1}.  Expand the form $\phi_i=y_i+p_i$, and reexpress the 
metric as an expansion 

\bqr 
{\cal K}= \sum b_{\omega}(p_i) \prod y_{\omega(i)} \ .  
\label{KahlerTwo}
\fqr  
The coefficients $b_{\omega(i)}(p_j)$ can be obtained, and they have a similar 
form as to $b_{\omega(i)}$ but with dependence on the integers $p_j$.  An 
important feature is that the solution to the 'metric' even in the presence 
of a set of integers $p_j$ that do not solve the algebraic equation can be 
found.  The metric around these points is not expected to be Ricci-flat, 
and it will not contain the required branch cut representing the resolvable 
singularity at the origin; in this case the origin is at $y_i=0$.  

Because the metric can be found at any possible sets of integers $p_j$, the 
presence of an actual integer solution can be determined by finding whether 
or not there is a branch cut at the origin about the sets of integers.  The 
analytic continuation of the sums in $b_{\omega(i)}(p_j)$ are required in 
order to determine this presence.  However, the explicit form of these 
coefficients is known due to the solution of the D-terms, which is found by 
the classical graph count \cite{Chalmers1}-\cite{Chalmers2}.  

The analytic solution to the metric can also be used to obtain the solution to 
sets of arbitrary algebraic equations, including those of Fermat. 

\vskip .2in 
\noindent{\it Analytic Continuation} 

There are various ways to analytically continue the infinite sum in the 
coordinates to find the branch cuts at the origin.  A direct sum is a bit 
problematic due to the complicated form of the coefficients $b_{\omega}(p_j)$, 
although this could be done.  The K\"ahler potential, and the metric, form 
about the origin can be obtained by setting all coordinates to $\phi_i=\phi$, 
with $\phi$ about the origin at $\phi=0$.  

This analytic continuation is simplified with the notation 

\bqr  
g= \sum a_n(p_i) x^n \ , 
\label{expansion}
\fqr 
with $x^n$ representing the terms $\phi^n$ after substituting $\phi_i=\phi$.  
The 
coefficients $a_n$ are found from the coefficients $b_\omega(p_i)$, which have 
been computed with the D-term solution.  

The same series in \rf{expansion} has the expansion in terms of logarithms 

\bqr 
g= \sum c_n(p_i) \ln^n(x) \ , 
\label{expansionlogs}  
\fqr 
which manifests the branch at the origin.  The coefficients $c_n$ can be found 
from those $a_n$.  Two explicit contour integrals around the origin will show 
whether or not there is a branch cut in the series \rf{expansion}.  

The coefficients $c_n$ in terms of $a_n$ are found by differentiation 
at $x=1$.  In terms of $x$, the K\"ahler potential derivatives evaluated 
at $x=1$ are, 
\bqr 
\partial_x^b g(x) = \sum a_n(p_i) {n!\over (n-b)!} \ ,  
\fqr 
and in terms of $\ln(x)$, the derivatives are, 

\bqr 
\partial_x^b g(\ln(x)) \vert_{x=1} =  \sum c_n (p_i) 
  \partial_x^b \ln^n(x)  \ . 
\fqr 
The identification 

\bqr 
\partial_x^b g(x) \vert_{x=1} = \partial_x^b g(\ln(x)) \vert_{x=1} \ , 
\fqr 
generates the identification of the coefficients.  
These identifications are found in closed form.  

The analytic continuation of the potential about the origin should  
have the form,  

\bqr 
g(x) = a x^\delta + \ldots \ ,  
\label{branch} 
\fqr 
which follows from the removable singularity occuring in the quotient 
description of the metric.  Multiple branch cuts at points $\vert x\vert 
\leq 1$ would not be physical in view of the quotient of the polynomial 
$Z_i \rightarrow G\cdot Z_i$, which describes the singularity.  The 
coefficients $a$ and $\delta$ can be found by two successive contour 
integrals around the origin.  

One contour integration follows from, 

\bqr 
I_1=\oint dx ~a x^\delta + \ldots = 
  \int_{\omega=0}^1 d\omega~ a e^{2\pi i\delta\omega} 
 = {a\over 2\pi i\delta} \left( e^{2\pi i\delta}-1\right) \ .   
\fqr 
\bqr
= {a\over 2\pi i\delta} \bigl( \cos(2\pi\delta)+i\sin(2\pi\delta)-1\bigr)
\label{integralone}
\fqr 
The real and imaginary parts of this integral generate the 
$a$ and $\delta$, involving the inversion of the sin and cos 
function.  All of the terms in the integrand which have 
integral powers of $x$ integrate to zero.   The solution to $a$ and 
$\delta$ follows from, 

\bqr 
{I_1^{\cal R}\over I_1^{\cal I}-1} = -\tan(2\pi\delta) \ ,
\fqr 
which can be used to find $\delta$.  Substituting this parameter into 
\rf{integralone} determines $a$.  Unfortunately, the inversion of a 
tan function is required, which slightly complicates the determination 
of $\delta$.  The determination of $I_1$ follows from integrating the 
logarithmic form of the potential with the same contour.   

A second contour integral is, 
\bqr 
I_2=\oint\oint dx~a x^\delta + \ldots = {a\over 2\pi i\delta} 
   \int_{\omega=0}^1 d\omega \left(e^{2\pi i\delta\omega}-1\right) \ , 
\fqr 
\bqr 
= {a\over 2\pi i\delta} \left( {1\over 2\pi i\delta}e^{2\pi i\delta} 
  -{1\over 2\pi i\delta} - 1 \right) \ . 
\fqr 
The remaining terms in the series integrate to, 

\bqr 
\oint d\omega ~a e^{2\pi i n \omega} = 0 \ , 
\fqr 
and 
\bqr 
\oint\oint d\omega ~a e^{2\pi i n\omega} = 
- {a\over 2\pi i n} \ . 
\fqr 
The second integral could be useful with further information of the 
analytic continuation.  

The determination of $I_1$ follows from the same contour integration 
as used to determine $a$ and $delta$.  The individual terms integrate 
as 

\bqr 
\oint d\omega ~\ln(e^{2\pi i\omega})^n = \int_0^1 d\omega 
~(2\pi i \omega)^n 
= (2\pi i)^n {1\over n+1} \ , 
\fqr 
and gives the form, 

\bqr 
I_1 = \sum_{n=0}^\infty c_n(p_i) {(2\pi i)^n\over n+1} \ .  
\fqr 
The evaluation of $I_1$ is simple, but the coefficients of $c_n(p_i)$ 
are found from the more complicated Calabi-Yau data.  These coefficients 
can be found from the classical graph count.  

\vskip .2in 
\noindent{\it Solution to Polynomials}  

Given the solution to the integral $I_1$, which leads to $a$ and $\delta$ 
through its real and imaginary parts, the counting of the solutions to the 
polynomials follows from the non-integrality of the parameter $\delta$.  
The function $\tan(2\pi\delta)$ vanishes whenever $\delta=n$; the vanishing 
of the function indicates the {\it non-presence} of the polynomial solution.  

The singularity in $\delta$ can be found from the algebraic curve.  With 
this value, the summation of the integers $p_i$ generates the allowed 
solutions to the polynomials; normalizing the $\tan(2\pi \delta)$ in the 
sum generates unity and a direct count of the integer solutions.  Also, 
the individual polynomial solutions are found by a non-vanishing of the 
number $\tan(2\pi \delta)$, which is unity after normalization.  A Heaviside 
step function would work also without normalization, which involves a Fourier 
transformation, but this is more complicated.  

The existence of a polynomial solution is found from 

\bqr 
C(p_i)=-\tan(2\pi\delta)^{-1} \arctan\Bigl(
  {I^{\cal R}(p_i)\over I^{\cal I}(p_i)-1}\Bigr) \ ,  
\label{solutioncount}
\fqr 
which is either one or zero, and $I=I_1$.  The complete sum, 

\bqr 
N= \sum_{p_i} C(p_i) \ , 
\label{totalcount}
\fqr 
generates the total number of solutions.  These functions $C(p_i)$ and 
$N$ depend on the curve, and are quite explicit due to the explicit form 
of the Calabi-Yau metric.  

\vskip .2in 
\noindent{\it Discussion} 

The explicit form of the Calab-Yau metrics permits a closed form solution 
to the existence of integer solutions to polynomials. This closed form 
requires some complicated sums, due to the form of the metric expanded 
around integer points.  

A counting function is given that allows the determination of the solutions.  
The sum over the integers generates the totality of these integer solutions 
to the polynomial equation $P(z)=0$, or systems of polynomial equations 
$P_i(z_j)=0$.  The formulae are quite explicit in terms of the metric data 
on the associated Calabi-Yau metric, expanded about integer points.  The 
well known example of Fermat's equations, or their generalizations, are an 
example.  The counting functions are derived from the explicit 
form of the metrics associated to the curves, with help from summations. 

\vfill\break 

\end{document}